\documentclass[preprint,11pt]{article}

\usepackage{bm}
\usepackage{epsfig}
\usepackage{color}
\usepackage{amssymb}
\usepackage{amsmath}

\newcommand {\lab}[1]{\label{eq:#1}}

\newcommand {\be}[1]{\begin{equation}{\lab{#1}}}
\newcommand {\ee}{\end{equation}}
\newcommand {\bea}{\begin{eqnarray}}
\newcommand {\eea}{\end{eqnarray}}

\begin{document}

\title{Phase Transitions in Models of Bird Flocking}

\author{
\textbf{H. Christodoulidi$^{1}$, K. van der Weele$^{1}$, Ch.G. Antonopoulos$^{2}$ and T. Bountis$^{1}$}\\
$^{1}$Department of Mathematics, Division of Applied Analysis  
and \\ Center for Research and Applications of Nonlinear 
Systems (CRANS),\\ University of Patras, GR-26500 Patras, Greece. \\
$^{2}$Institute for Complex Systems and Mathematical Biology (ICSMB), \\
 Department of Physics, University of Aberdeen, AB24 3UE Aberdeen, United Kingdom}


\maketitle
\textbf {We dedicate this paper to the loving memory of Professor John S. Nicolis. His contributions to nonlinear science and its applications to biological information processing, his erudite presentations and devotion to his students and colleagues will always be an inspiration to us all.}

\begin{abstract}
The aim of the present paper is to elucidate the transition from collective to random behavior exhibited by various mathematical models of bird flocking. In particular, we compare Vicsek's model [Viscek {\it et al.}, Phys. Rev. Lett. {\bf 75}, 1226 -- 1229 (1995)] with one based on topological considerations. The latter model is found to exhibit a first order phase transition from flocking to decoherence, as the ``noise parameter''  of the problem is increased, whereas Viscek's model gives a second order transition. Refining the topological model in such a way that birds are influenced mostly by the birds in front of them, less by the ones at their sides and not at all by those behind them (because they do not see them), we find a behavior that lies in between the two models. Finally, we propose a novel mechanism for preserving the flock's cohesion, without imposing artificial boundary conditions or attracting forces. 
\end{abstract}

                             

\section{Introduction}
Collective animal behavior is an active field of research, where motile particles self-organize without the presence of a leader, constituting a prime example of a living {\it complex system}. Everybody is familiar with flocks of starlings (Sturnus Vulgaris) which display an impressive cohesion and synchronization in their movements. Fundamental questions arising naturally in such systems concern the microscopic rules of interaction between the individuals, which result in the organized dynamic structures on a macroscopic scale. The ultimate goal is to find appropriate simple models that describe the observed behavior in a satisfactory way.

Even though it is a relatively young area of research, collective animal behavior has already 
generated a vast literature, where many ideas and models have been proposed, corresponding
to a variety of interesting approaches to the problem. Our perspective in the present study
is motivated by the search for a minimal model, in the spirit of that introduced by Vicsek {\it et al}.\cite{vics95} The Vicsek model describes the movement of active particles on the plane, where at each step, the direction in which the particle moves is decomposed into one averaged over the directions of all others inside a range of radius $r$ plus a small shift in a random direction. So, the main idea governing the {\em local} dynamics is simply this: Each particle is influenced by its nearby neighbors but also exercises to some extent its own individual motivation or ``free will''.
  
Several other models have since been proposed, which introduce modifications to the above interaction rule, constituting a class of so--called `Vicsek type models'. Toner and Tu \cite{tu95} developed a coarse--grained dynamical description, which is in accordance with the original Vicsek model, see Ramaswamy \cite{Rama} and Viscek {\it et al.} \cite{vics12} for a detailed review. Gr{\'e}goire {\it et al.} \cite{chate03,chate04} focused on the nature of phase transitions of Vicsek--like models, adding vectorial noise (instead of angular), thus complementing the results of Vicsek {\it et al}. \cite{vics95} Also other types of models have been proposed in the literature. For example, Garc{\'i}a Cant{\' u} Ros and co-workers \cite{anton} consider angular displacements governed by one--dimensional deterministic maps and show for different initial conditions that collective motion is attracted to the onset of chaos where flocking phenomena are dominant. An interesting type of model can also be found in the work of Hildenbrandt {\it et al.} \cite{hilden}, which has a rather complicated formulation but yields very realistic simulations of bird flocks.

A breakthrough in the research on flocking was a sequence of experiments in Rome, performed by a large group of scientists who used modern equipment to observe the movement of starlings and especially the interactions between neighbors within the group. This ambitious venture, known as the StarFlag project, provided many valuable insights \cite{starflag1,starflag3}. One of the most fundamental results was that each bird interacts with maximally 6 or 7 neighbors; and since this number was found to be independent of the mutual distances between the birds, the researchers named this type of interaction {\it topological} \cite{parisi}. 

This new type of interaction differs from that of the Vicsek model, or any other {\em metric} model, in which the interactions are limited to a specified range of radius $r$ around each bird. In the metric models the number of interacting birds may be constantly varying, whereas in topological models the group can expand and contract without altering the number of individuals communicating at every time step.

Our aim in the present paper is to compare the flocking properties of the standard Vicsek paradigm to those of topological models. In Section \ref{models} we describe Vicsek's model and its topological counterpart, and pay particular attention to the important concept of the {\em flocking index} $v_{\alpha}$, being the modulus of the average velocity of the birds. This index will serve as our order parameter. It varies between unity (complete alignment) and zero (absence of alignment), depending on the level of ``free will'' in the model (expressed by the parameter $\eta$). For the Vicsek model $v_{\alpha}$ has been claimed (in the limit when the number of birds $N$ goes to infinity) to exhibit a second order phase transition from 1 to 0 as $\eta$ is increased \cite{vics95}. By contrast, in the topological model we find evidence that $v_{\alpha}$ undergoes a {\it first order} transition.   

Next, in Section \ref{vrange}, we modify the topological model to take into account the visual field of the birds. That is to say, they are influenced mainly by those in front, less by those on their sides and not at all by the ones behind them. With this modification, the transition of the flocking index from 1 to 0 (as $\eta$ grows) lies in between the above two cases. It still resembles a first order transition, as for the standard topological model, but the slope of $v_{\alpha}$ is less steep. A further improvement to the model is proposed in Section \ref{nobound}, where we introduce a simple mechanism to preserve the cohesion within the flock. Unlike in previous models, the proposed mechanism does not resort to heuristic attracting forces or unrealistic periodic boundary conditions. Instead, we incorporate the natural instinct of birds which tells them, whenever they stray beyond the outer edges of the flock, to steer back toward its center. Finally, in Section \ref{Concl} we draw our main conclusions.

\section{Vicsek and Topological Type Models \label{models}} 

Both the Vicsek and the topological model can be formulated in terms of a map on the plane: 
\begin{eqnarray}\label{map}
\vec {r}_i(t+\Delta t) = \vec {r}_i(t) + \vec {v}_i(t) \Delta t, ~~ i=1,\ldots,N, ~\vec {r}_i, \vec {v}_i \in \mathbb{R}^2, 
\end{eqnarray}  
where $N$ is the number of particles, $\vec {r}_i$ identifies the position of the $i$--th particle on the plane and $\vec {v}_i $ is its velocity, which is of constant modulus $v$. $\vec{v}_i(t)$ is updated at each time step as follows: 
\begin{equation}\label{vel}
\vec {v}_i(t+\Delta t) = 
\left( \begin{array}{cc}
\cos \vartheta_i(t)  & -\sin \vartheta_i(t)    \\
\sin \vartheta_i(t)  & \cos \vartheta_i(t) 
\end{array} \right)
\cdot \vec {v}_i(t), ~~~ i=1,\ldots,N,
\end{equation}  
i.e. through the one--parametric rotation matrix. At this point, we need to define the dynamics of the angles $\vartheta_i$, reflecting the rules of communication among the individuals inside the flock\footnote{The angle inside the Vicsek model\cite{vics95} represents the angle  with respect to the $x$--axis, instead here $\vartheta_i$ is taken with respect to the velocity vector. Vicsek's law and equation 
(\ref{ang1}) are equivalent.}.

In the standard Vicsek model, the angles are at each time step updated by the map
\begin{eqnarray}\label{ang1}
\vartheta _i(t+\Delta t) = <\vartheta_i (t)>_r   + \eta_i(t)  ~~~ i=1,\ldots,N
\end{eqnarray} 
where $\eta_i(t)$ is a ``noise'' term taken randomly (at each time step again) from the uniform distribution $[-\eta/2, \eta/2]$, 
and $ <\vartheta _i(t)>_r$ is the average angle of all particles inside a disk of radius $r$ with the particle $i$ at its center.  Instead, in the topological model the angles are given by 
\begin{eqnarray}\label{ang2}
\vartheta _i(t+\Delta t) = <\vartheta_i (t)>_n   + \eta_i(t)  ~~~ i=1,\ldots,N
\end{eqnarray} 
where $<\vartheta _i(t)>_n$ is the average over the angles of the $n$ nearest interacting neighbors  
of the particle $i$. Particle $i$ is included in both averages of equations (\ref{ang1})
and (\ref{ang2}).
 
There are fundamental differences between the two models, as mentioned also in Ginelli {\it et al.}\cite{ginelli}. The most important difference is that Vicsek's model is ``metric'', in the sense that the number of interacting particles within each  disk depends on the radius $r$. Instead, in the topological model two individuals experience the same interaction, independently of their distance, and this property implies stronger cohesion of the group.

There are also similarities between the two models. For instance, in both cases the particles tend to disperse away from the ``center of mass''. To avoid  this, periodic boundary conditions are often assumed in the numerical simulations (see e.g. Vicsek {\it et al.}\cite{vics95}), so that the group moves inside a square of side length $L$ and therefore has constant density $\rho = N/ L^2$. If one regards the boundaries as open, on the other hand, the particles tend to drift away and their density rapidly decreases, making it necessary to impose additional assumptions to obtain stable results. In the papers of Gr{\'e}goire {\it et al.}\cite{chate03,chate04} this was called the problem of {\em zero--density limit} and attracting forces were introduced at long distances to preserve group cohesion.

\section{Flocking Index and Phase Transitions}

In this section, we begin by describing a flocking index $v_{\alpha}$ that we will use as the {\em order parameter} monitoring phase transitions in the models under study. This quantity is the modulus of the average velocity, providing an estimate of the extent of alignment of the birds' velocities within the group. Following Vicsek \cite{vics95} we define: 
\begin{eqnarray}\label{fi}
v_{\alpha } = \frac{1}{N v} \big| \sum_{i=1}^N \vec {v}_i \big| ~~~.
\end{eqnarray} 
This parameter varies from zero (no average alignment of the velocities) to unity, when all particles move in the same direction. It is important to mention that the suitability of this index is related to the kind of collective motion we want to study. In models of fish schools, where the fish tend to form big circles and rotate around them, the above index would be zero despite the fact that the motion is completely coherent. The same would be true when the flock expands and contracts periodically about a fixed point in space. In such cases, one has to apply another type of index. One example of an alternative index was introduced by Garc{\'i}a Cant{\' u} Ros {\it et al.} \cite{anton}. It would be desirable to have a universal flocking index that can recognize various types of coherence, but in the present work we will not pursue this point and simply work with the classic flocking index of equation (\ref{fi}).

\begin{figure}
\centering
\includegraphics[scale=0.30 ]{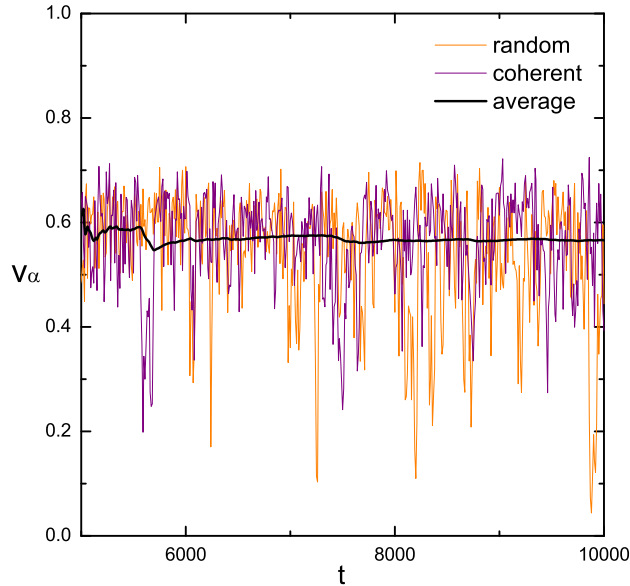}
\caption{  The temporal evolution of the velocity modulus $v_{\alpha }(t)$ given in Eq.(\ref{fi}) 
for random (orange) and coherent (purple) initial conditions, together with its running time average (black). The latter, which is the time-average from $t = 5000$ to the current $t$-value, is seen to converge to the value $0.57$.  \label{ave}  }
\end{figure}

The value of the flocking index was calculated numerically in our examples  as follows: We first compute the quantity $v_{\alpha }$ of equation (\ref{fi}) as a function of time, integrating the map (\ref{map}). Ignoring then an initial time window, we evaluate its running time average, which exhibits much smaller fluctuations and converges better to a certain value between 0 and 1. In Fig.\ref{ave} we show the instantaneous and time averaged $v_{\alpha } (t)$ for a particular example of the topological model. Now, since the model has a random part,
we average also over 7 different runs, in order to get the final value for the flocking index.  
Throughout the paper the modulus of the velocity is set to $v=0.03$, as in Vicsek {\it et al.} \cite{vics95}. Initially, all particles are randomly distributed inside a square with density $\rho =4$.

\begin{figure}
\centering
\includegraphics[scale=0.20 ]{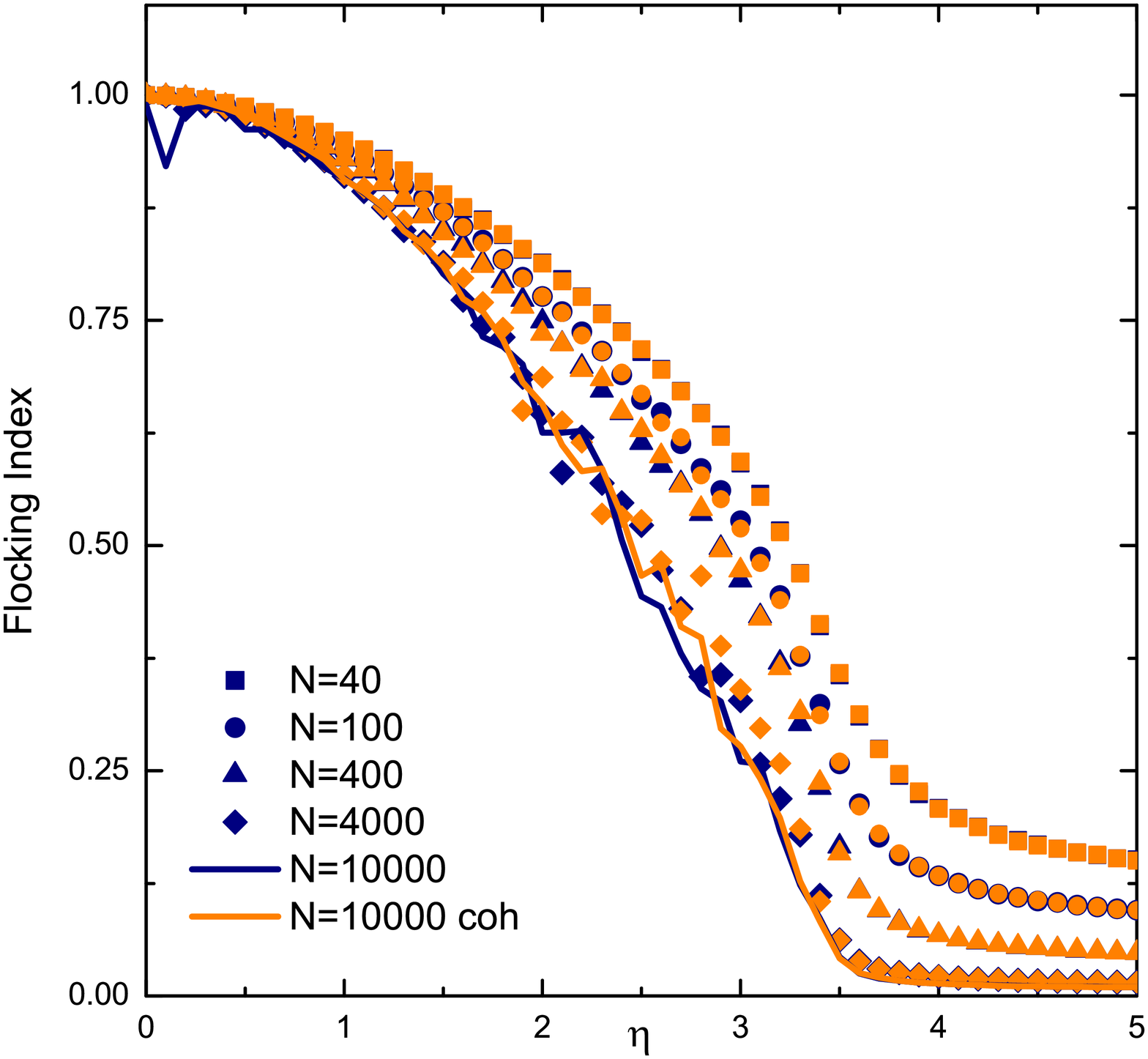}
\includegraphics[scale=0.20 ]{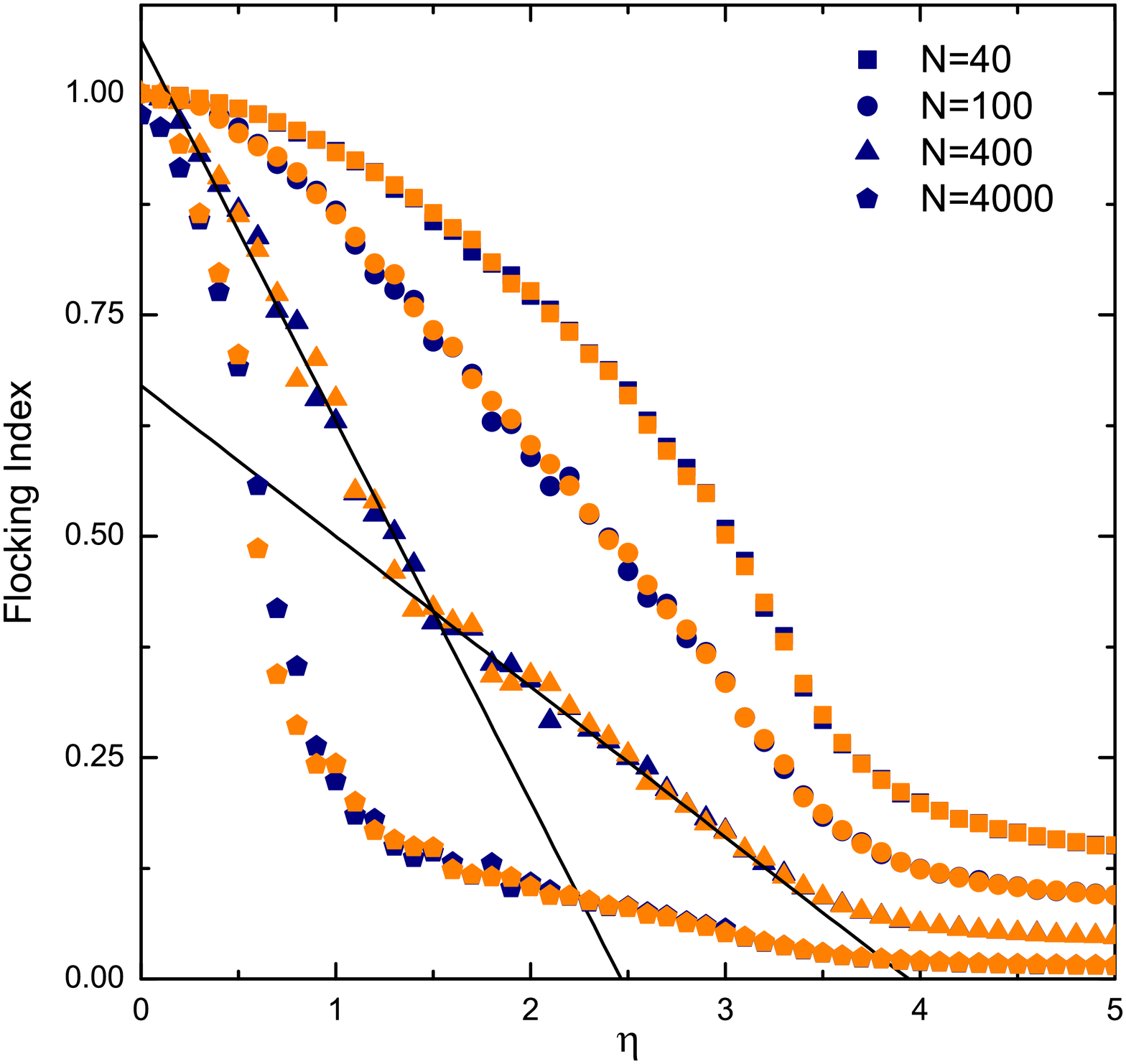} 
\caption{ The flocking index as a function of the noise level $\eta$ for (a) Vicsek's model and (b) the topological model with $n=7$ interacting neighbors. Both panels show the flocking index for 
different system sizes $N$. With blue we represent the random initial conditions 
and with orange the coherent ones. The two straight lines in the right plot are guides to the eye, illustrating the sudden jump of slope in the curve of the flocking index for $N=400$. \label{fi1}  }
\end{figure}

 \begin{figure}
\centering
\includegraphics[scale=0.19 ]{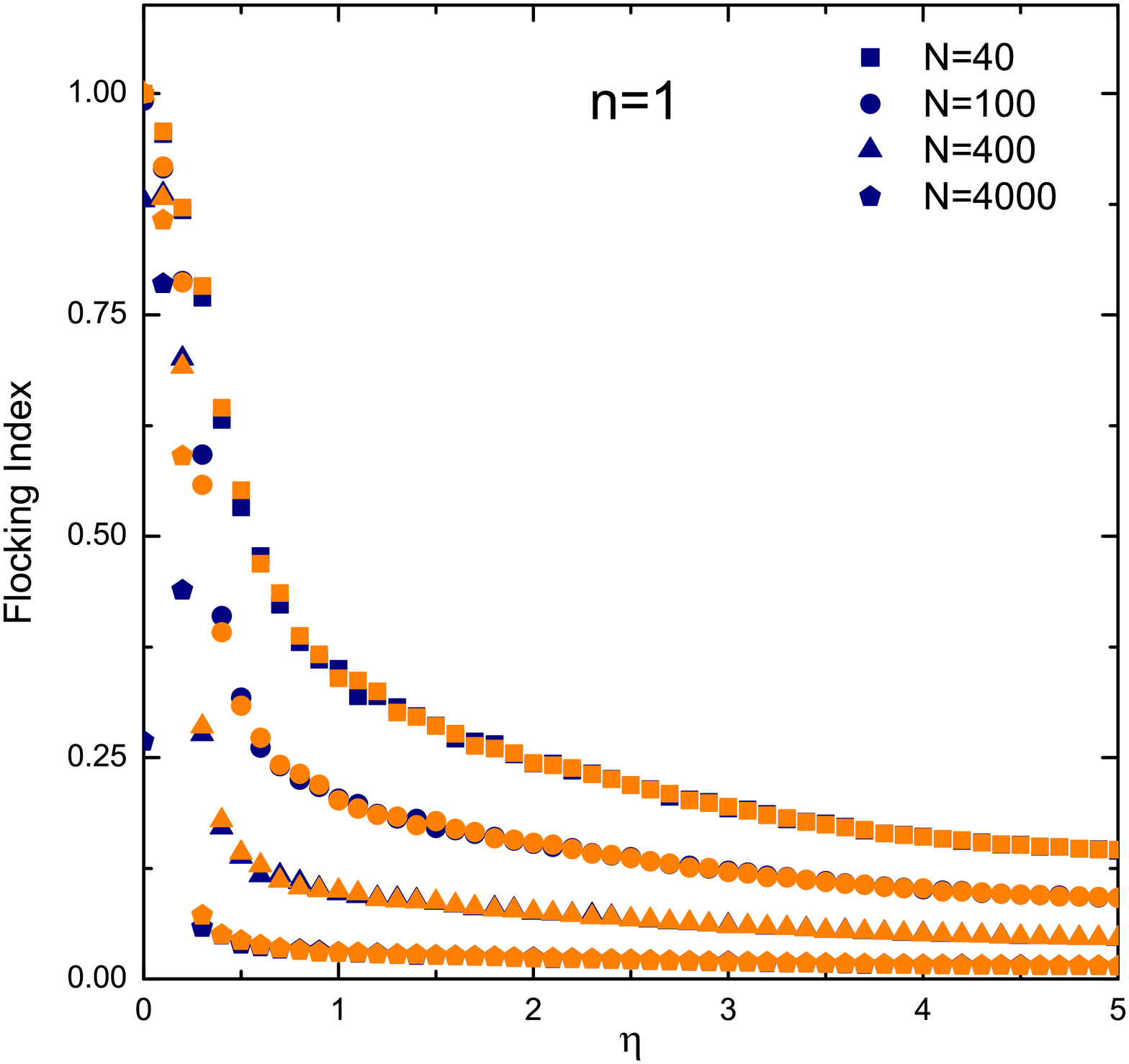}
\includegraphics[scale=0.19 ]{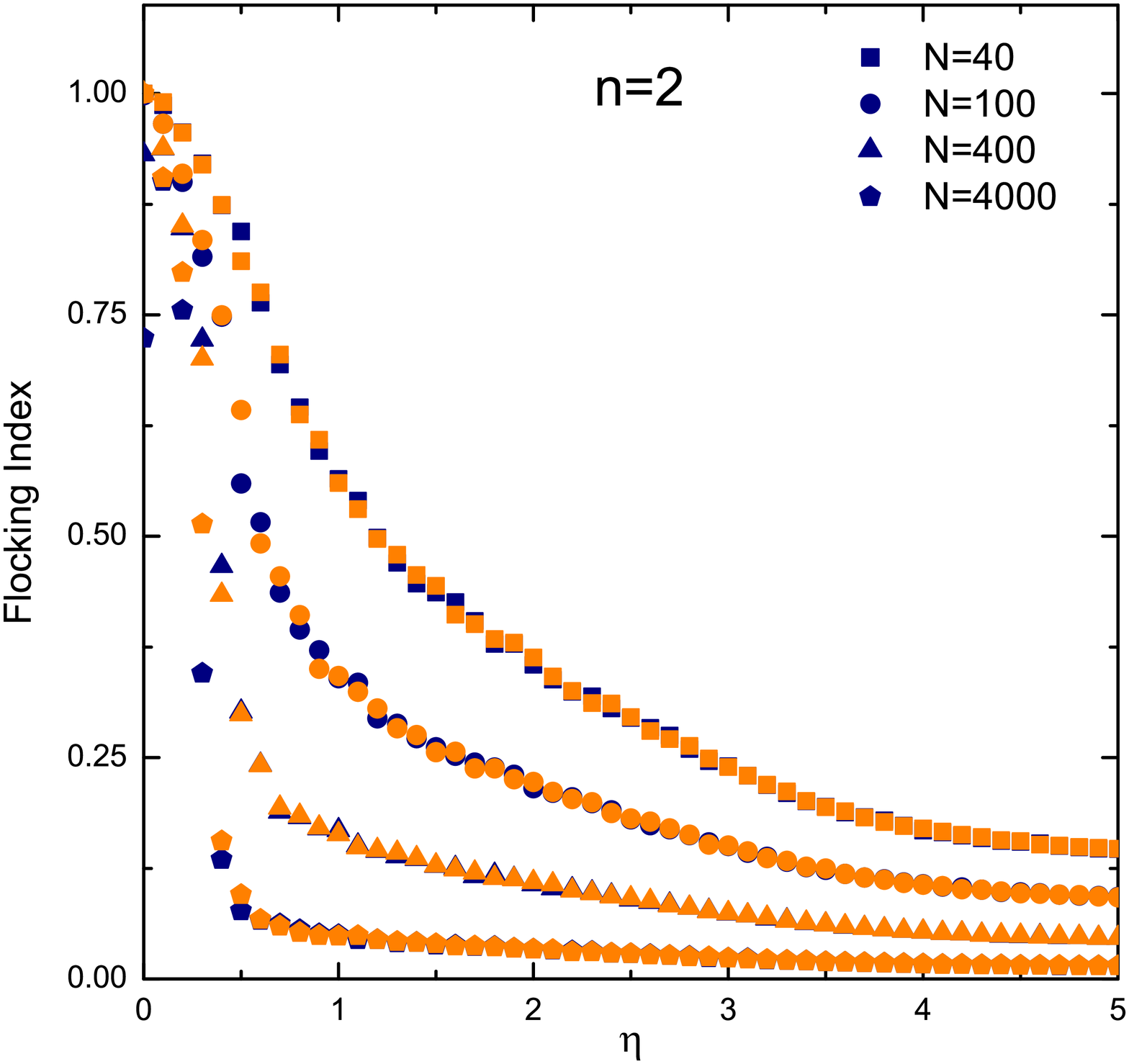} 
\includegraphics[scale=0.19 ]{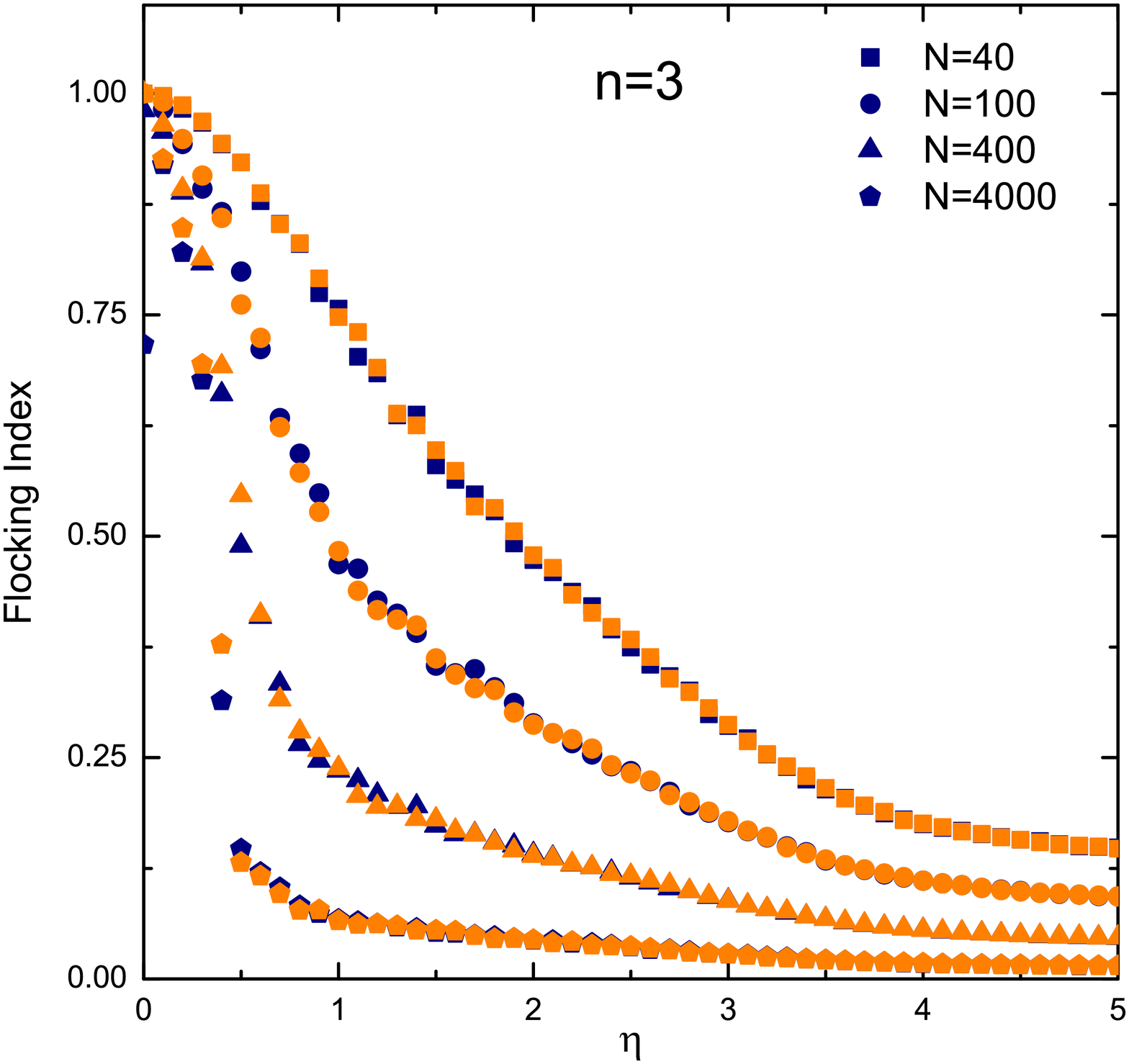}
\includegraphics[scale=0.19 ]{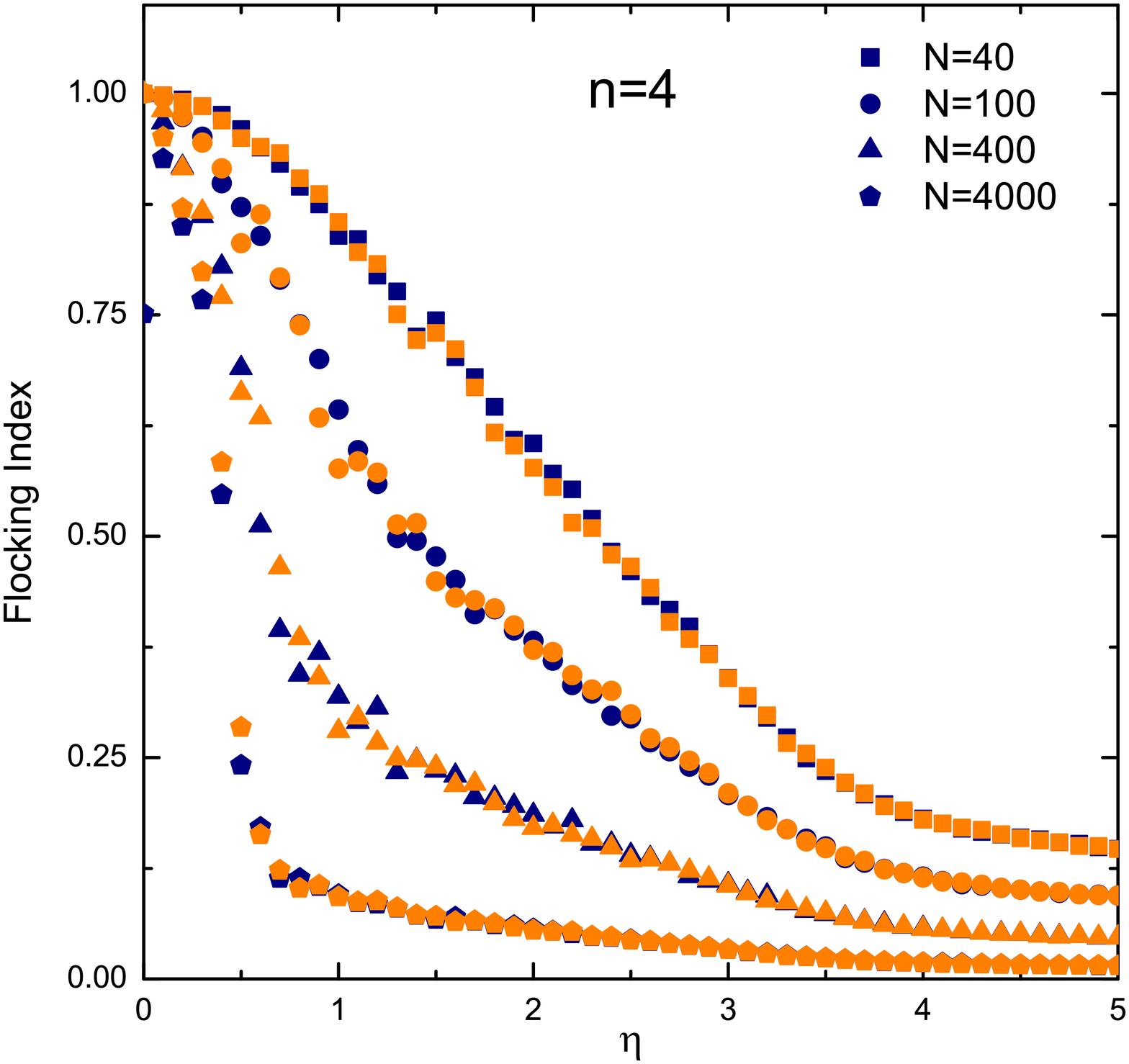}
\includegraphics[scale=0.19 ]{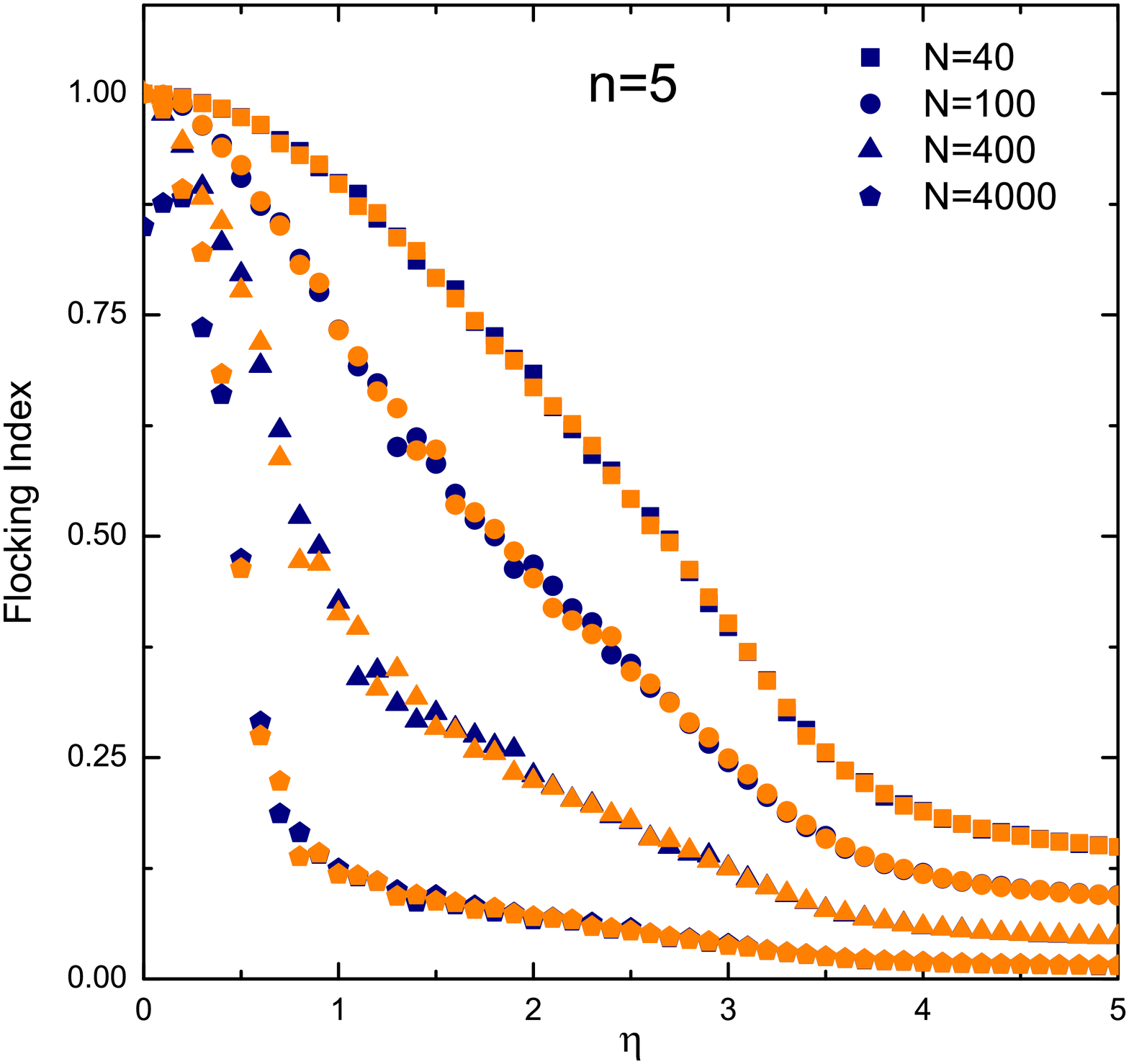}
\includegraphics[scale=0.19 ]{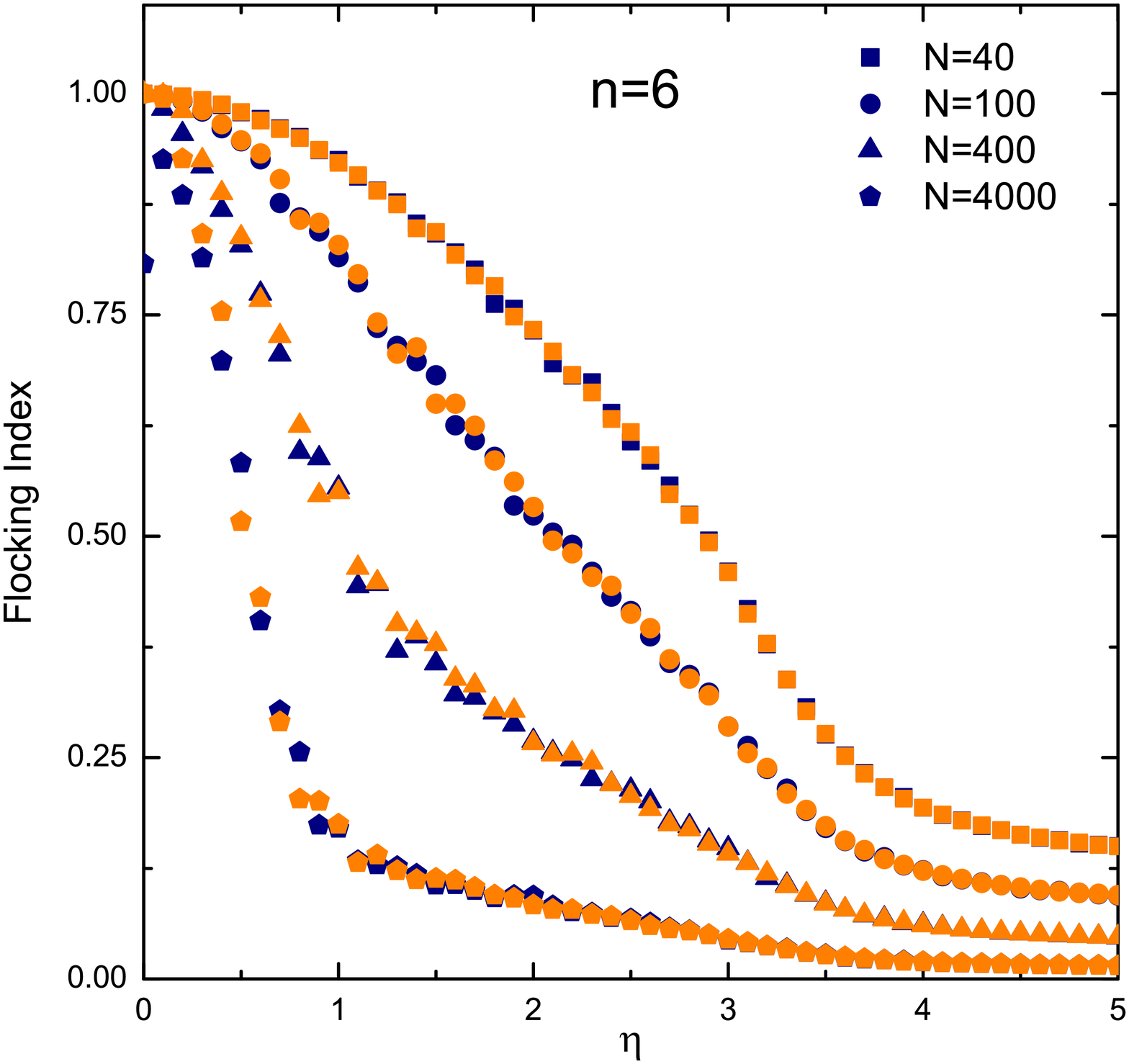}
\caption{ Flocking index versus the noise level $\eta$ for the topological model
(see also Fig. \ref{fi1}(b)) for $n=1$
to $n=6$ interacting neighbors respectively. Evidently the curves appear to 
attain a similar form for $n\geq 3$. \label{fi2}}
\end{figure}
 
Following previous studies of this problem, we now raise the value of the noise amplitude $\eta$ in (\ref{ang1}) and (\ref{ang2}) and plot the flocking index as a function of $\eta$. Each experiment was repeated for two different initial conditions: (i) When the particles start out with random angles in the range $[-\eta/2, \eta/2]$ (random initial condition) and (ii) when all velocities initially point in the same direction (coherent initial condition). The results for the Vicsek model and the topological model with $n=7$ neighbors are plotted side by side in Fig. \ref{fi1}. We  observe that the models give very similar results when the number of birds is small ($N=40$) but become increasingly different for growing $N$.

As expected, Fig. \ref{fi1}(a) shows that the Vicsek model has the tendency to stabilize to a concave universal curve in the limit of large $N$. For $N=40$ or $N=100$ finite size effects are still present but these gradually disappear as $N$ is increased. The order parameter $v_{\alpha}$ is a continuous function of $\eta$ and undergoes a second order phase transition from 1 to 0, approximated by the function
\begin{equation}
\label{critical}
v_{\alpha}(\eta) \sim (1-\eta/\eta_{cr})^{\beta}
\end{equation}
where $\eta_{cr} \approx 3.5$ and $\beta$ is a non-integer positive critical exponent. In fact, Vicsek and co-workers in \cite{vics95} found that $\beta=9/2$. The different colors of the data points indicate different initial conditions (random and coherent, respectively) and we see that the results are independent of the initial condition.

By contrast, the results of the topological model in Fig. \ref{fi1}(b) show a different behavior. The curve of the flocking index ceases to be concave beyond a certain value of $N$ ($N_0 \approx 100$) and becomes convex. When $N$ is increased further, we witness the formation of a conspicuous kink, where the slope of the curve makes a sudden jump. For $N=400$ this jump lies around $\eta = 1.5$ (two straight lines have been drawn to guide the eye, illustrating the sudden change of slope). The results for growing $N$ suggest that the flocking index may become discontinuous in the limit $N \rightarrow  \infty$, making a sudden jump from 1 to 0 (at $\eta=0$) which is the hallmark of a first order phase transition. Just as in the case of the Vicsek model, the results are found to be independent of the initial conditions.

In Fig. \ref{fi2} (a)--(f) we investigate the influence of the number of interacting neighbors in the topological model. Above we had chosen $n=7$, which is the experimentally observed value in three dimensions \cite{parisi} but it is appropriate (since we here work in the two-dimensional plane) to also study smaller values. The plots of Fig. \ref{fi2} show that the behavior of the flocking index changes gradually when $n$ is raised from 1 to 7. In particular, the discontinuous jump at $\eta=0$ (in the limit $N \rightarrow  \infty$) becomes very evident for small $n$. Simultaneously, the concave part (which may be taken as a sign of a second order transition) is increasingly absent for small values of $n$. We conclude that the nature of the phase transition becomes increasingly first-order as $n$ is decreased.

This may be explained by the fact that (as $n$ becomes small) the communication between the birds is reduced, which means that small noise levels are sufficient to break the coherence of the group. When two or three birds have the possibility to act as an isolated entity they may easily start to behave independently (and even break away) from the rest of the group. The coherence of the group is naturally more robust for higher
$n$ values. Nevertheless, when we take Figs. \ref{fi2}(a-f) at face value and assume that they {\it all} show a sudden jump from 1 to 0 at $\eta=0$ in the limit $N \rightarrow  \infty$, we arrive at a surprising conclusion: The observed coherence in bird flocks is (for any fixed value of $n$) a {\it finite-size effect} which would disappear when the group size $N$ became infinitely large.

\section{A Model Involving Visual Range Interactions \label{vrange}}

\begin{figure}
\centering
\includegraphics[scale=0.35 ]{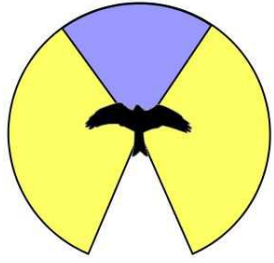}
\caption{ The visual field of a starling is divided in three regions: the binocular (blue),
the monocular (yellow) and the visionless area (blank). \label{eye}  }
\end{figure}

In this section we introduce a model that is on the one hand, based on the simple idea of
the original Vicsek model and on the other hand similar to the topological model.
To be specific, in the Visual Range Interactions (VRI) model each 
starling tries to align with a fixed number of neighbors by a simple linear law
under the presense of noise, without the introduction of forces, 
taking into account the natural visual field of each bird.
It is known that the vision of birds is divided into three main areas: the binocular, 
the monocular and the visionless area (Fig. \ref{eye}) in such a way that different importance 
is given to the neighbors in each area.

Martin {\it et al.}\cite{eye} gives a detailed description of starling's vision, with precise angles 
for each of the fields. These angles are relative to the position of the eye. 
To fix a value for these angles,
let us assume that the eyes are pointing in the normal position. In this case, 
the bird's binocular field of vision corresponds to 21 degrees in the front, 
the monocular to 143 degrees for each eye (to the sides) 
and 53 to the visionless area (see Fig. \ref{eye}). 

These visual fields affect directly the nature of interactions, since
a bird tends to follow more the ones in {\it front} than those on the side and does not notice at all 
the birds in the back. In addition, the observation of the StarFlag group on starling flocks \cite{feder} states that 
`there is higher probability in finding its nearest neighbor on the side, rather than
in front or back along the direction of motion'. At first sight, one might infer from
this statement that birds tend to interact more intensively with their side neighbors.
Thinking more  carefully however, one can see that this need not be true. For example, car drivers tend to 
pay a lot more attention to the cars in front of them, ones with which collisions are much more likely, than the side ones.

Let the movements of the birds in two spacial dimensions be given by the map 
(\ref{map}), where the velocity of each bird $\bar{v}_i $ is updated by
the relation (\ref{vel}). As stated in Section \ref{models}, the crucial question
is how to define the dynamics for the angle $\vartheta _i$. In order to take into account
the visual range, we express the influence of the flock on a bird by a weighted average
over $n$ nearby birds. In particular, the updated angle is given by 
\begin{eqnarray}\label{vis}
\vartheta _i(t+\Delta t) = <\vartheta _i(t)>^{vis}_n + \eta_i(t)  ~~~ i=1,\ldots,N
\end{eqnarray}
where $\eta_i(t)$ is the noise inside the interval $[-\eta / 2, \eta / 2 ]$ and  
\begin{eqnarray}\label{avevis}
<\vartheta _i(t)>^{vis}_n  = \frac{ 1}{c_1 n_1 + c_2 (1-n_1)} \cdot \big( c_1 \sum_{j=1}^{n_1} \vartheta _j(t) + c_2 \sum_{j=n_1+1}^{n} \vartheta _j(t) \big)
\end{eqnarray}
the weighted average with $c_1 > c_2$ the field--weights. 
The first sum in the expression (\ref{avevis}) corresponds
to the average over  $n_1\leq n$  angles of the birds that lie inside the binocular field of vision of the $i$--th bird
and the second sum is for the birds inside its monocular vision, while we completely disregard the birds in
the backside, no matter what their distance. In this way, the first sum is dominant and a priority to the front birds is given. 

\begin{figure}
\centering
\includegraphics[scale=0.19 ]{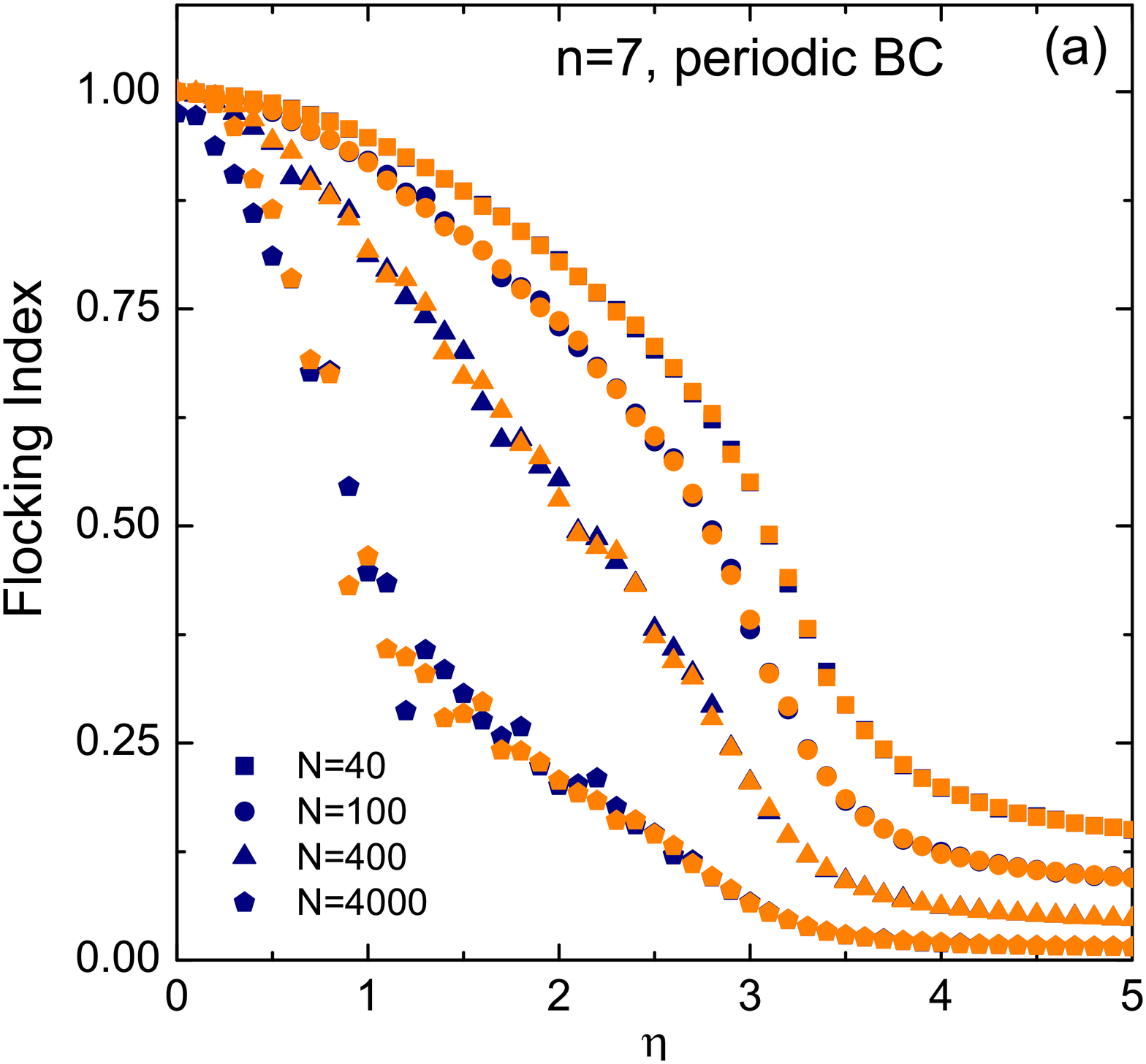}
\includegraphics[scale=0.19 ]{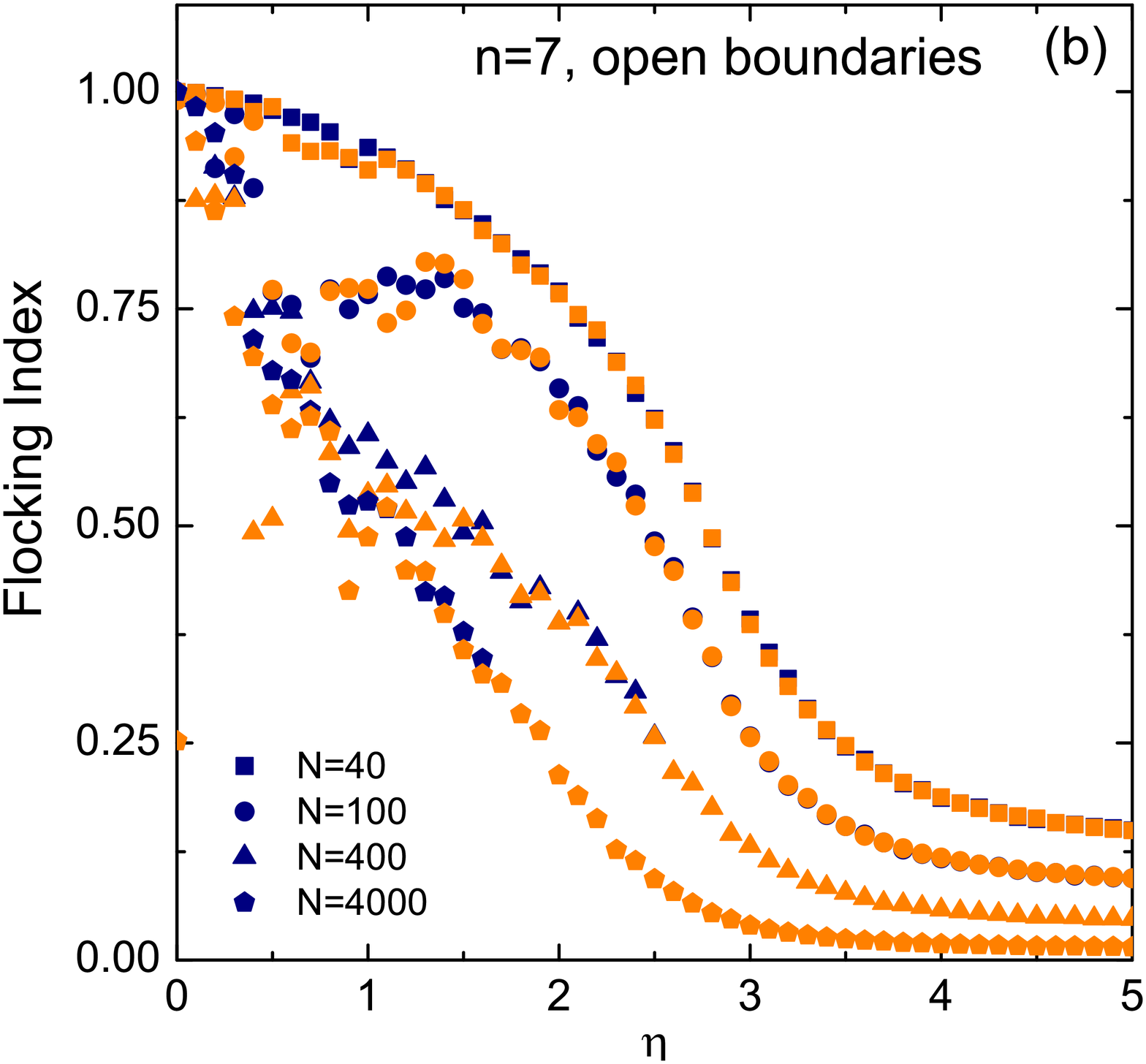}
\caption{ Flocking index for $n=7$ neighbors versus the noise strength $\eta$ for 
different group sizes $N$ of the VRI model with (a) periodic
and (b) open boundary conditions. The blue data points represent 
the random initial conditions and the orange points the coherent ones. \label{vri}  }
\end{figure}

We study the phase transitions for this model. Figure \ref{vri}(a) shows the Flocking Index versus the noise strength
of the VRI model for $n=7$ neighbors with field--weights $c_1=3$, $c_2=1$. Although less steep than the topological model, 
its phase transition remains of first order. To settle this point conclusively, we are currently studying this issue 
in the light of the modern classification of phase transitions in small systems \cite{borrmann}. We hope to report on this 
in a future publication.

\section{Visual Range Interactions without boundaries: A mechanism for preserving group cohesion\label{nobound}} 

\begin{figure}
\centering
\includegraphics[scale=0.29 ]{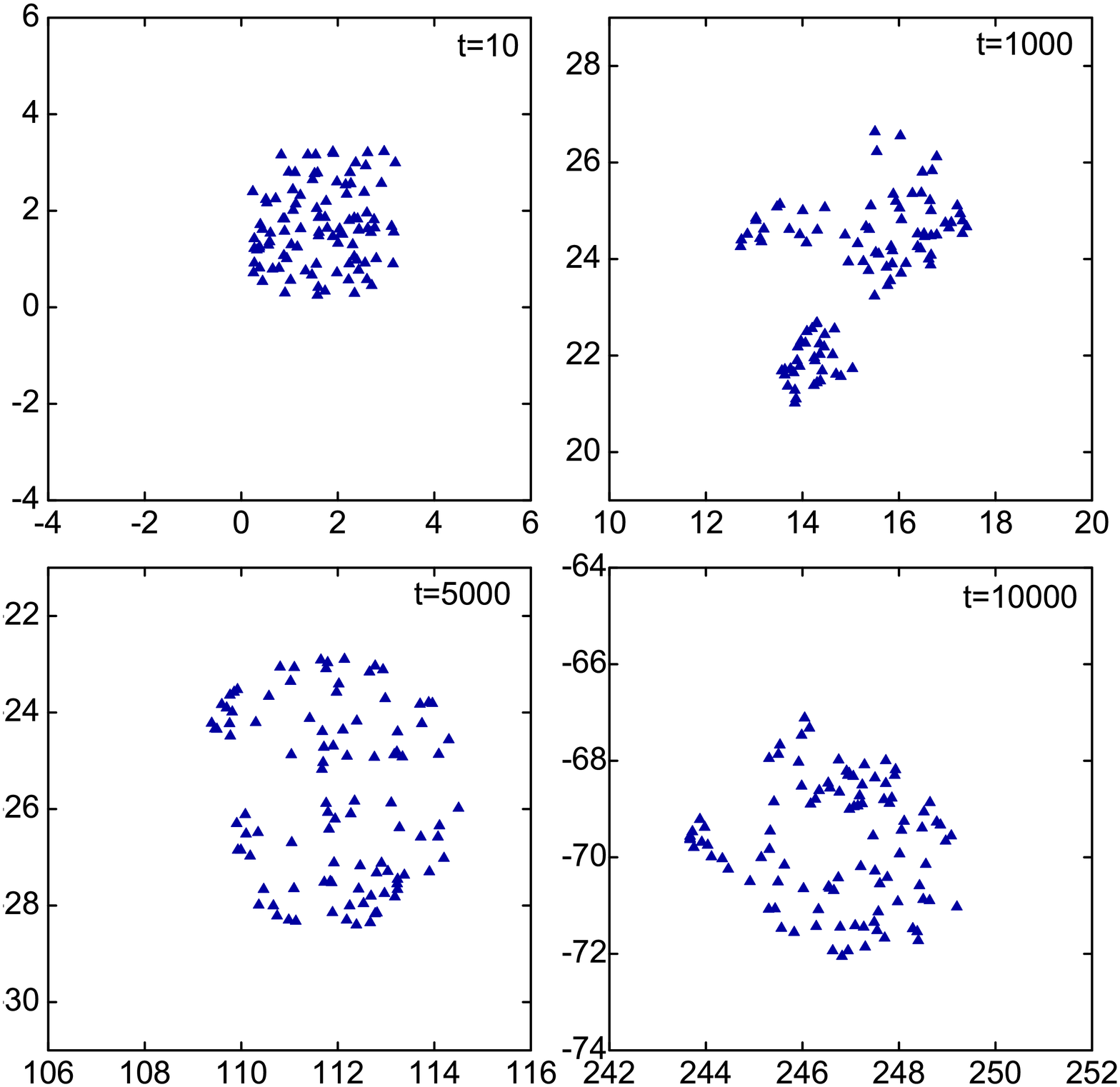}
\caption{ Snapshots showing the evolution of a bird flock consisting of $N=100$ birds, computed with the Visual Range Interactions model.  \label{vri2}  }
\end{figure} 

Cohesion problems arise in the direct simulations of the VRI model defined in (\ref{vis}) in the absence of any boundary conditions. As in Vicsek's and the topological model, free boundaries result in splitting the flock into several parts that diffuse in space, leading to the so--called zero density limit.
To avoid this problem, additional assumptions have been proposed. For example, in many works \cite{chate03,chate04,giardina} attracting forces are added, acting when the distance of a bird from the rest becomes larger than
a threshold range $r_a$, which we will refer to as {\it range of recall}. In the present paper, to avoid the introduction of such forces, we simply assume that, when a bird exceeds
a prefixed distance $r_a$ from the center of mass of the flock, 
the alignment law is replaced by a {\it recall law}, i.e. relation (\ref{vis}) 
is replaced by a rotation of the velocity vector $\bar {v}_i$ of the $i$--th bird
towards the center of mass of the flock. Consequently, the velocity modulus
of the bird does not change and in this way the bird catches up with the rest by moving straightforward towards the coherent flock, without spending time in turns or maneuvers before getting back inside the group.

Numerical simulations for different values of the parameters, like the threshold range $r_a$, the strength of noise and the field--weights, produce some interesting results. For example, focusing on the value of the recall range parameter $r_a$, within which birds tend to align while outside they move straight towards the center of the flock, we distinguish three distinct cases: i) $r_a<R$, $R$ being the radius of the flock, ii) $r_a=R$ and iii) $r_a>R$.

In the first case we observe that the behavior of the flock remains coherent and the birds cluster 
in small groups that rotate around the center of mass, without any escapes from the flock, provided that the low noise level is not too high. Such behavior is reminiscent of analogous results found in Garc{\'i}a Cant{\' u} Ros {\it et al.} \cite{anton}. Consequently, the choice of a small range restricts the birds to move mostly towards the center and thus the recalling law dominates the alignment law. On the other hand, the flock appears more homogeneous when the range of recall is equal to the flock's radius. This behavior seems to be more realistic, since the birds move in more synchronized fashion, changing direction simultaneously. Now, if the range $r_a$ is larger, the flock typically splits into two or three parts, which follow the same direction of motion and eventually merge back together, re-establishing the initial flock. Additionally, the phenomenon of expanding and contracting of the flock is also present.

By contrast, when the noise strength is high, the birds move randomly inside the flock and mix rapidly, while the whole group is essentially wondering in space (the center of mass of the flock remains trapped inside a relatively small area). This behavior corresponds to a flock that is not well organized, with its components moving as if they were in panic under the presence of a possible external threat. 

Figure \ref{vri2} shows four snapshots of a simulation of the VRI model without boundary conditions at different moments in time. We have chosen the recall range to be equal to the radius $R=3$ of the flock, for $N=100$ particles, $\eta = 0.4$ and weights $c_1=20$, $c_2=1$ (strong dominance of the binocular field). At early times, $t=10$, the birds are still highly concentrated. At later times, see the plots at $t=1000, 5000$ and $10000$, the flock changes shape and expands or contracts a little. However, at all times cohesion remains strongly present and no bird escapes. By looking at the numbers along the axes one can follow how the whole flock wanders in space. 

\section{Conclusions \label{Concl}} 

In this work, we have studied the flocking properties of different models describing the behavior of birds flying in large groups in a two--dimensional plane. We compared the standard Vicsek (metric) paradigm, where each bird has a fixed radius of influence, to the so--called topological models, where each bird has a fixed number of neighbors which whom it interacts. In particular, we examined the {\em flocking index}, which varies between unity (complete alignment) and zero (when the velocities point equally in all directions). In the original Vicsek model, this index is known to exhibit a second order phase transition from 1 to 0 as the noise (or ``free will'') parameter $\eta$ is increased. However, for topological models we find evidence that the transition is of first order, since the flocking index (as a function of $\eta$) does not appear to converge to a smooth curve as the number of birds $N$ increases.

We have also checked the phase transition for the metric model by Garc{\'i}a Cant{\' u} Ros {\it et al.} \cite{anton} (to which we added a noise term $\eta(t)$ in the same way as in the Vicsek model and the topological model) and found a second-order transition. This further strengthens the observation that metric flocking models typically exhibit a second-order transition.

We also modified the topological model to take into account the visual field of the birds, which means that each individual bird is influenced most by those in front of it, less by those on either side, and not at all by those behind it. What we have found is that the transition of the flocking index from 1 to 0 (as $\eta$ grows) is now less steep than in the original topological model, but still resembles a first order transition.

In our models we have achieved cohesion of the group at all times (a well-known problem in the modelling of flocks) without resorting to heuristic attracting forces or unrealistic boundary conditions, by introducing the ``recall mechanism'' which represents the natural tendency of birds to steer back toward the center of the flock whenever they wander beyond its outer edges.

\paragraph{Acknowledgments}
The authors wish to thank  A. Ponno for insightful and helpful discussions on the various types of models that are used to describe bird flocking.
Part of this work was supported by the European research project ``Complex Matter'', funded under the ERA-NET Complexity Program.  This research has been co-financed by the European Union (European Social Fund – ESF) and Greek national funds through the Operational Program "Education and Lifelong Learning" of the National Strategic Reference Framework (NSRF) - Research Funding Program: Thales. Investing in knowledge society through the European Social Fund. The numerical simulations were performed at the TURING cluster of the University of Patras.

\bibliographystyle{elsarticle-num}

\end{document}